\documentclass[conference]{IEEEtran}
\usepackage{cite}
\usepackage{amsmath,amssymb,amsfonts, bbm}
\usepackage{algorithmic, graphicx}
\usepackage{xcolor}
\usepackage{tabularx}
\usepackage[ruled,vlined]{algorithm2e}
\usepackage{soul} % Load the soul package
\sethlcolor{yellow} % Set the highlighting color to green
\usepackage{flushend}
\def\BibTeX{{\rm B\kern-.05em{\sc i\kern-.025em b}\kern-.08em
    T\kern-.1667em\lower.7ex\hbox{E}\kern-.125emX}}
\def\BibTeX{{\rm B\kern-.05em{\sc i\kern-.025em b}\kern-.08em T\kern-.1667em\lower.7ex\hbox{E}\kern-.125emX}}

\DeclareMathOperator*{\clip}{clip}

\title{DeepMPR: Enhancing Opportunistic Routing in Wireless Networks through Multi-Agent Deep Reinforcement Learning}
\author{Saeed Kaviani, Bo Ryu, Ejaz Ahmed, Deokseong Kim, Jae Kim, Carrie Spiker, Blake Harnden \\
Episys Science Inc, Poway, CA \qquad and \qquad The Boeing Company, Seattle, WA  \\
\small{\texttt{\{saeed, boryu, ejaz, david\}@episci.com}} \\
\small{\texttt{\{jae.h.kim, carrie.v.spiker, blake.j.harnden\}@boeing.com}} 
}
\begin{document}
\maketitle
\thispagestyle{plain}
\pagestyle{plain}

\begin{abstract}
Opportunistic routing relies on the broadcast capability of wireless networks. It brings higher reliability and robustness in highly dynamic and/or severe environments such as mobile or vehicular ad-hoc networks (MANETs/VANETs). To reduce the cost of broadcast, multicast routing schemes use the connected dominating set (CDS) or multi-point relaying (MPR) set to decrease the network overhead and hence, their selection algorithms are critical. Common MPR selection algorithms are heuristic, rely on coordination between nodes, need high computational power for large networks, and are difficult to tune for network uncertainties. In this paper, we use multi-agent deep reinforcement learning to design a novel MPR multicast routing technique, DeepMPR, which is outperforming the OLSR MPR selection algorithm while it does not require MPR announcement messages from the neighbors. Our evaluation results demonstrate the performance gains of our trained DeepMPR multicast forwarding policy compared to other popular techniques.
\end{abstract}

% --------------------------------------------------------
\section{Introduction}
In recent years, wireless networks have witnessed significant advancements, leading to their widespread adoption across various applications and domains. These networks are now expected to deliver higher bandwidth, extensive coverage, and enhanced reliability. However, traditional wireless networking paradigms fall short in meeting the escalating demands for increased connectivity, robustness, and quality of service (QoS).  

In this paper, we focus on opportunistic routing (OR), also called \textit{anypath routing}. OR is a routing technique that capitalize on the inherent broadcast nature of the the wireless medium \cite{biswas2004opportunistic, biswas2005exor}. Unlike traditional routing algorithms that rely solely on a single, predetermined path from the source to the destination, OR enables the utilization of multiple paths to increase the reliability and efficiency of data transmission. Notably its effectiveness becomes particularly pronounced in environments characterized by high mobility or frequent link failures, such as mobile ad-hoc networks (MANETs) and vehicular ad-hoc networks (VANETs).

The majority of the OR techniques rely on a flooding mechanism in dynamic networks, which can incur substantial cost. Therefore, intelligent path selection becomes imperative to mitigate these expenses. Furthermore, frequent changes in network topology and network uncertainties pose additional challenges as any routing protocol deployed in such scenarios trend to augment processing overhead due to control messages, often necessitating the use of flooding mechanism. This is where machine learning techniques can offer substantial benefits by predicting network dynamics and accounting for uncertainties. By doing so, excessive network flooding can be reduced, along with the need for saturating the network with frequent control messages. 

In this work, we leverage the advancements in deep learning and adopt a multi-agent deep reinforcement learning (MA-DRL) approach to discover a multicast forwarding deep neural network (DNN) policy. This policy enables individual nodes to make informed decisions regarding the forwarding of multicast packets, thereby reducing unnecessary flooding overhead and maximizing the overall efficiency of the network, especially in high data rate scenarios where network congestion is prevalent.

% -----------------------------------------------------------
\subsection{Related Works}
% SMF -> OLSR unicast TC -> MPR 
In multicast routing where a node communicates to all (or many) nodes in the network, OR is deployed and optimized for efficient use of the flooding technique. This is particularly performed within the simplified multicast forwarding (SMF) protocol for MANETs \cite{macker2012simplified}. To reduce the spectral cost of broadcast, SMF and other popular schemes typically employ the connected dominating set (CDS) \cite{hu2004connected} where local topology information decreases the number of packet retransmission and the network overhead. A similar approach is used in a popular MANET routing protocol, the optimized link state routing (OLSR) \cite{clausen2003optimized} during its network discovery process. OLSR protocol is a proactive unicast routing where each packet is intended to be sent to a single destination. It maintains a complete topology map of the network and calculates the shortest path between nodes for the unicast packets. Prior to unicast routing decisions, OLSR collects network topology information via OR flooding mechanism to distribute traffic control (TC) messages. The OLSR TC messages are multicast packets. To reduce TC multicast forwarding overhead, OLSR uses the \textit{multi-point relaying (MPR)} concept \cite{qayyum2002multipoint}. The MPR set is a subset of neighbor nodes that are responsible to forward the multicast packet. Particularly, in OLSR the MPR set is selected to be a subset of its 1-hop neighbors that guarantees to reach all of the 2-hop neighbors. Each node selects its MPR set independently. The nodes that are selected by at least one of their neighbors can only generate and forward multicast TC messages. TC messages are flooded to the entire network as part of the control traffic. Therefore, the selection of the MPR set is critical to reducing the total number of TC messages flooded into the network. TC messages are used to construct a global view of the network topology and to compute the shortest path routes to all destinations. The selection of MPRs is crucial for the performance of OLSR, and several algorithms have been proposed to optimize this process. 
 
\begin{figure}[t]
\centering
\includegraphics[width=6.5cm]{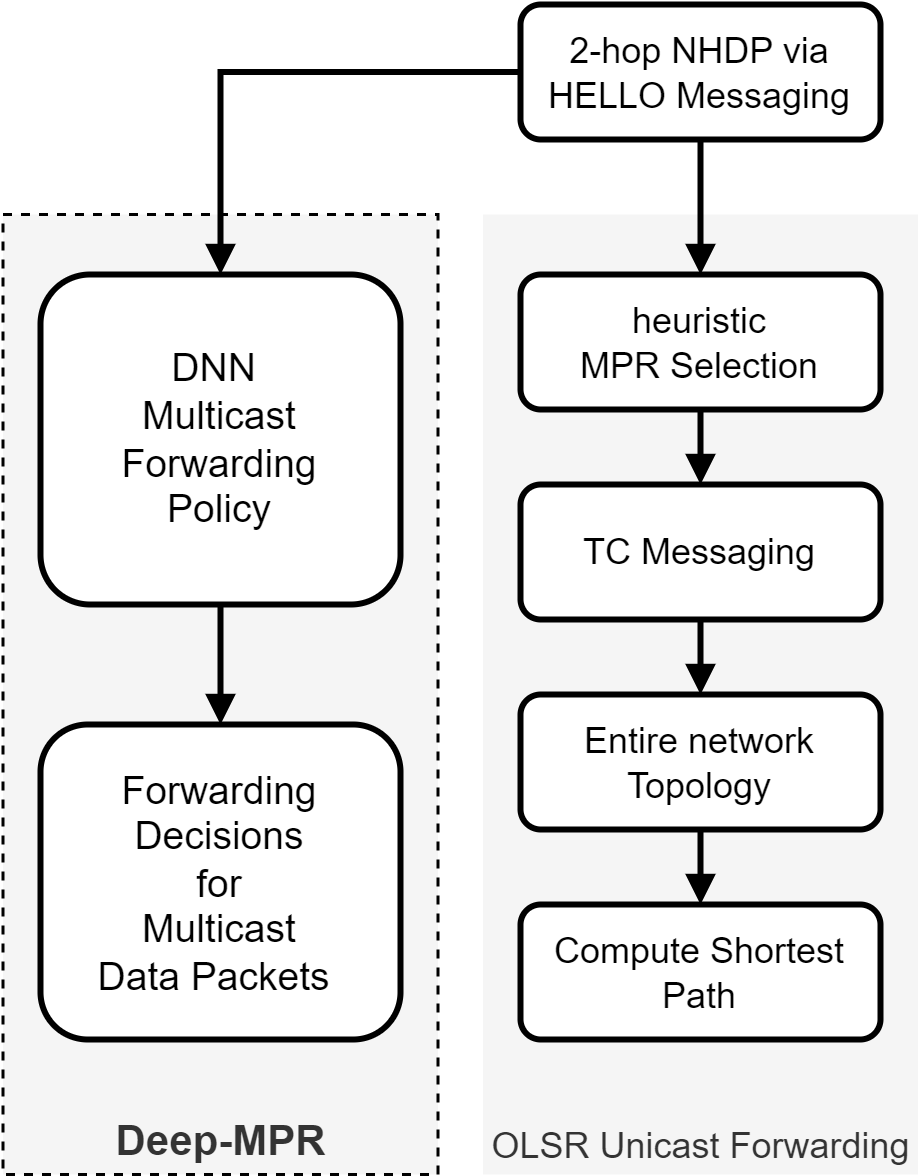}
\caption{Deep-MPR is compared to the MPR used in OLSR unicast forwarding.}
\label{fig:deepmpr_vs_olsr}
\end{figure}

In this work, we use the MPR concept used in the TC messaging mechanism of the OLSR and apply it for the multicast data packet forwarding. When OLSR or similar protocols are extended for multicast forwarding, there will not be any TC messages as the data packets are intended to reach the entire network. We use MPR set for the flooding of the data packets rather than the TC messages and propose the Deep-MPR multicast forwarding policy using DRL. This is summarized visually in Fig. \ref{fig:deepmpr_vs_olsr}. The MPR set selection is even more crucial in multicast forwarding protocols as it reduces the main multicast data forwarding overhead. However, the selection of the smallest MPR set for a node is known to be an NP-hard problem \cite{garey1979computers}, and OLSR adopts a heuristic algorithm that finds MPR set with a size within a bound from local minimum. Note that the minimization of the size of the MPR set does not necessarily lead to the minimization of the overhead in the network. 

The problem of global minimization of the MPR set size has been formulated as an integer linear programming (ILP) that can be solved efficiently for relatively small or medium-size networks \cite{maccari2018where}. The MPR set selection algorithm is formulated as ILP in  \cite{maccari2018where,gantsou2009revisiting} or as genetic algorithm \cite{singh2019evolutionary}. However, the global minimum is not practical as it requires centralized coordination, and high computational power for large networks, and therefore not scalable. Moreover, it does not directly minimize the overhead of the network, and is difficult to trade off between overhead and delivery ratio when it is possible. Hence, our approach is to use DRL to design a DNN policy to select an MPR set efficiently, i.e. DeepMPR, and improve the network overhead significantly. 

Routing involves successive selections of the nodes in the network leading to a combinatorial number of possibilities; therefore the globally optimal route via exhaustive search is computationally prohibitive in large networks. This is even more highlighted for opportunistic routing problems where the decision possibilities include multicast forwarding. The proposed decentralized discrete optimization approaches in network routing are mostly heuristic \cite{chen1999distributed}. The sequential decision-making nature of the routing problems fits the Markov decision process (MDP) \cite{bellman1957markovian} making it a prime candidate to apply reinforcement learning (RL). 
DRL algorithms are successfully applied to complex high-dimensional problems, mainly due to the use of DNN for function approximations \cite{mnih2015human}. DRL algorithms previously applied in mobile ad-hoc networks (MANETs) for the routing optimization \cite{yu2018drom, stampa2017deep, cui2021scalable}. In the context of opportunistic routing, DRL is also applied to the multicast packets but given the unicast data flows \cite{kaviani2021deepcq+}.

% ===========================================================
\section{Multicast Routing in Ad-Hoc Network}
% -----------------------------------------------------------
\subsection{Network Model}
Consider a wireless ad-hoc network supporting multicast data flows incoming from all the nodes intended to reach the rest of the network, and routed through multiple hops. Our design assumes that the MANET is highly dynamic, which is a challenging environment that requires highly robust, efficient, and scalable routing protocols. For the sake of training and testing, we considered the Gauss-Markov MANET mobility model. The Gauss-Markov mobility model \cite{liang1999predictive} has been shown to solve both of these problems \cite{Ariyakhajorn2006}. 

Each network is modeled as a directed graph $\mathcal G:=(\mathcal V, \mathcal E)$ with $N$ nodes or vertices denoted by $\mathcal V$ and the edges or links in between denoted by $\mathcal E$. We consider multicast packets denoted by $p = (s, i, h)$ where it is currently in the forwarding queue of node $i$, generated at a source node $s$, and it is forwarded from previous hop $h$ to node $i$. $\mathcal N_1(i)$ and $\mathcal N_2(i)$ are one-hop and two-hop neighbor sets of node $i$, respectively, and are obtained during the two-hop neighborhood discovery process using \texttt{HELLO} messaging. We also define $\mathcal{\widetilde{N}}_1(z;i) = \mathcal N_1(i) \cap \mathcal N_1(z)$ for any $z \in \mathcal N_2(i)$ which includes all $\mathcal N_1(i)$ neighbors which are directly connected to $z$. Similarly, we define $\mathcal{\widetilde{N}}_2(y; i) = \mathcal N_2(i) \cap (\mathcal N_1(y) \cup \mathcal N_2(y))$ as $\mathcal N_2$ neighbors reachable by node $y \in \mathcal N_2(i)$. 
The MPR set $\mathcal M(i)$ is a subset of the neighbors of node $i$ selected as MPR candidates. $\mathcal K(i)$ is the set of all neighbors of $i$ that have selected node $i$ as their MPR candidates. The global true network link (adjacency) matrix is denoted by $\mathbf L \in \{0,1\}^{N \times N}$. Note that only partial link matrix knowledge is available at each node $i$ and it is denoted by $\widehat{\mathbf{L}}_i \in \{0,1\}^{n_i \times n_i}$ where $n_i = |\mathcal N_1(i) \cup \mathcal N_2(i)| + 1$ is the number of nodes in the neighborhood of node $i$.

% To formulate the problem mathematically, we assume sequential propagation of the multicast packets, where $\mathcal X_k$ is the set of nodes that received packet $p$ for the first time at time $k$, 

% For a packet $p$ only $M(s)$ will forward it, $N_1(s)$ will receive the packet, if not duplicate, all $u \in N_1(s)$ will forward it, 
% $|N_1(s)| + \sum_{u \in M(s)} |N_1(u) - N_1(s)|$
% At step $k$, set of all nodes have received packet $p$ are $F_k = \union_{f \in F_{k-1} \union 
% $H_k$ = all nodes have already received $p$
% $H_k - H_{k-1}$ = nodes that received $p$ at time $k$
% \begin{align}
% \mathcal X_0 = \{s\}, & \quad \mathcal H_0 = \{s\} \\
% \mathcal X_1 = \mathcal N_1(s) , & \quad \mathcal H_1 = \mathcal M(s) \\
% \mathcal X_2 = \bigcup_{u \in \mathcal H_1} \mathcal N_1(u) - \mathcal X_1 \cup \mathcal X_0, & \quad \mathcal H_2 = \bigcup_{u \in \mathcal H_1} \mathcal M(u)\\
% \cdots \\
% \mathcal X_{k+1} = \bigcup_{u \in \mathcal H_{k}} \mathcal N_1(u) - \bigcup_{l=1}^k \mathcal X_l& \quad \mathcal H_{k+1} = \bigcup_{u \in \mathcal H_{k}} \mathcal M(u)
% \end{align}

\subsection{Multicast Routing Optimization Problem}
We aim to maximize the overall goodput in the entire network. The goodput is aggregated across all multicast destinations and only considers the non-duplicate packets received. In this paper, we consider all mobile users to act as multicast receivers. Aggregate goodput, $Gp(bps)$, is the primary performance metric to compare various SMF protocols. It is defined as the number of non-duplicate bits per unit of time (or bits per second (bps)) received at all destinations (or received at all group nodes if multicast). We also consider the overhead ratio as the ratio of the number of bits per unit of time throughput (all received packets), $Thr$, and the aggregate goodput and it is given by 
\begin{equation}
    OH = \frac{Gp(bps)}{Thr(bps)}
\end{equation}
The aggregate goodput is one measure of end-to-end performance when it is measured over increasing traffic loads. Hence, we compare the DeepMPR results under both mobility and increased traffic loading. The more intelligent selection of MPR reduces the overhead of the network, and as the traffic load increases results in higher aggregate goodput. 

In Algorithm \ref{alg:mpr_algorithms}, the popular MPR and CDS-based multicast forwarding protocol are summarized using the heuristic MPR selection aglorithm and their forwarding decisions compared versus the Deep-MPR. The OLSR routing protocol uses the S-MPR (i.e. source-specific MPR) forwarding algorithm, which uses the heuristic algorithm given the two-hop neighborhood information. Subsequently, SMF nodes select and announce a subset of their bi-connected one-hop neighbors as MPR nodes to ensure flooding coverage to all two-hop neighbors. To forward a packet, an S-MPR node needs to receive a unique multicast packet from any of its bi-connected neighbors, and the neighbor from which the packet was received has to select the node as an MPR. Therefore, S-MPR requires duplicate detection, MPR selection signaling, and packet previous-hop knowledge. The NS-MPR (i.e. non-source specific MPR) algorithm is a variant of S-MPR that combines source-specific MPRs to form a single CDS, which forwards every unique packet if and only if its selector list is non-empty.

\begin{algorithm}
\SetAlgoLined
$v$: The node that performs the computation.\\
% $N_1$: 1-hop neighbor set of $v$.\\
% $N_2$: 2-hop neighbor set of $v$ excluding $v$ and $N_1(v)$.\\
% $\mathcal{\widetilde{N}}_1(z) = \mathcal N_1(v) \cap \mathcal N_1(z)$: $N_1$ neighbors directly connected to $z \in N_2$.\\
% $\mathcal{\widetile{N}}_2(y) = \mathcal N_2(v) \cap \mathcal N_$: $N_2$ neighbors reachable by node $y$.\\
% $M(v)$: $N_1$ neighbors selected as MPRs by $v$.\\
% $K(u)$: The nodes selected $u$ as MPR (MPR-Selectors).\\
% $\mathsf{RtrPri}$: The router priority, we use $|N_1|$ as a router's priority, break ties with the address.\\
\SetKwProg{Fn}{Function}{:}{}
\Fn{\texttt{Select\_MPR()}}{
    \While{$\mathcal{N}_2 \neq \emptyset$}{
        Find $\mathcal{\widetilde{N}}_1(z)$ for all $z \in \mathcal N_2$\;
        Find $\mathcal{\widetilde{N}}_2(y)$ for all $y \in \mathcal N_1$\;
        \For{$z \in \mathcal N_2$ where $|\mathcal{\widetilde{N}}_1(z)| = 1$}{
            Select nodes of $\mathcal{\widetilde{N}}_1(z)$ as MPR candidates\;
        }
        Select $y \in \mathcal N_2$ with the largest $|\mathcal{\widetilde{N}}_2(y)|$ as MPR\;
        \For{$u$ selected as MPR}{
            Remove $u$ from $\mathcal N_1$\;
        }
        \For{$y \in \mathcal{\widetilde{N}}_2(u)$}{
            Remove $y$ from $\mathcal N_2$\;
        }
    }
}

\Fn{\texttt{announce\_MPR()}}{
    \For{each node $v \in \mathcal V$}{
        Share MPRs $\mathcal M(v)$ with neighbors $\mathcal N_1$\;

    \If{$v \in \mathcal M(u)$}{
        Add $u$ to its MPR-Selectors $\mathcal K(v)$, i.e., $\mathcal K(v) \leftarrow \mathcal K(v) \cup \{u\}$\;
    }
    }
}
\Fn{\texttt{Forward()}}{
    \If{\texttt{S-MPR}}{
        Forward all multicast packets if received from its MPR-Selectors\;
    }
    \If{\texttt{NS-MPR}}{
        Forward multicast packets if $\mathcal K(v) \neq \emptyset$\;
    }
    \If{\texttt{Deep-MPR}}{
        Forward based on the stochastic Deep-MPR DNN Policy }
    %     \If{$v$ has the highest address in $N_1(v)$ and MPR-Selectors}{
    %         Sort nodes in $M(v)$ based on the router priority expression $RtrPri$\;
    %         \Return{$M(v)$}\;
    %     }
    % }
    % \If{\texttt{MPR-CDS}}{
    %     \If{$v$ has the highest address in $N_1(v)$ and MPR-Selectors}{
    %         Sort nodes in $M(v)$ based on the router priority expression $RtrPri$\;
    %         \Return{$M(v)$}\;
    %     }
    % }
    % \If{\texttt{E-CDS}}{
    %     \If{$\mathsf{RtrPri}(v) = \max_{u \in N_1 \cup N_2}\mathsf{RtrPri}(u)$}{
    %         Add itself $v$ as MPR; $M(v) \leftarrow M(v) \cup \{v\}$\;
    %     \If{$w_{max} = \argmax_{w \in N_1} \mathsf{RtrPri}(w)$}{
    %         If no path exist from $w_{max}$ to the other nodes in $N_1 \cup N_2$ with higher $\mathsf{RtrPri}$ than $v$, select $v$ as forwarder.
    %     }
    %     }
    % }
}
\caption{MPR-based Multicast Forwarding Algorithms}
\label{alg:mpr_algorithms}
\end{algorithm}

% ===========================================================
\section{Reinforcement Learning Approach for Multicast Routing}
\begin{figure}[t]
\centering
\includegraphics[width=8.5cm]{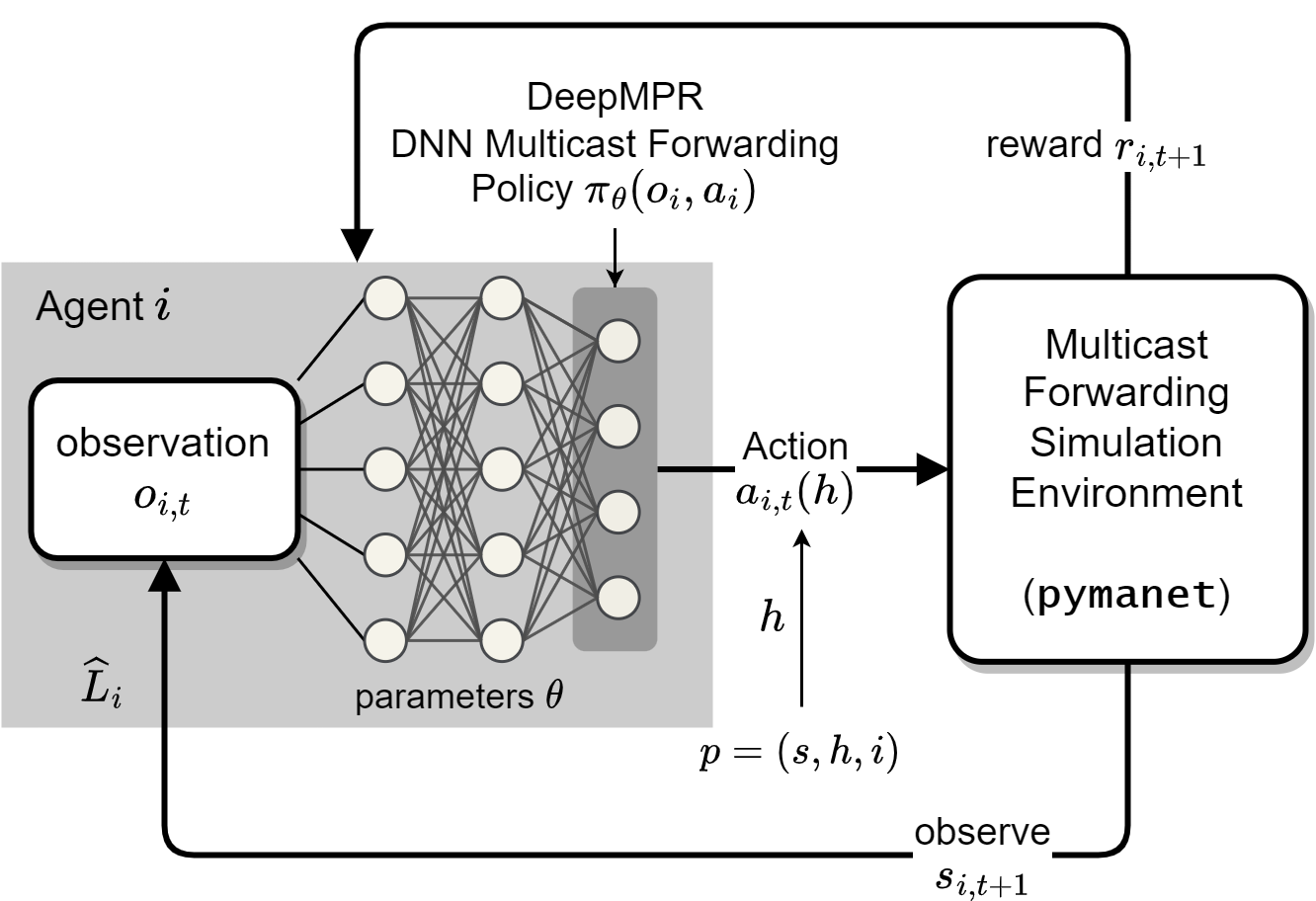}
\caption{The RL framework for the multicast routing policy design.}
\label{fig:drl_framework}
\end{figure}

Network routing is a sequential decision-making problem where the decisions are made by the nodes in the network given local knowledge of the network conditions. Hence, it fits as a multi-agent \textit{decentralized partially-observable Markov decision process (Dec-POMDP)} \cite{oliehoek2016concise}. Dec-POMDP tries to model and optimize the behavior of the agents while considering the environment's and other agents' uncertainties. The POMDP is represented by $(\mathcal{S};\mathcal{O};\mathcal A;\mathcal T;\mathcal R; \gamma)$ with a set of states $\mathcal{S}$ describing the possible configurations of the agent(s), a set of observations $\mathcal{O}$, a set of actions $\mathcal{A}$, transition dynamics $\mathcal T$), reward function $\mathcal R$, and discount factor $\gamma$. The actions are selected using a stochastic policy $\pi_\theta: \mathcal{O}\times\mathcal{A} \rightarrow [0,1]^{|\mathcal N_1(i)|}$ (parameterized by $\theta$), which results in the next state defined by the environment's state transition function $\mathcal{T}: \mathcal{S} \times \mathcal{A} \rightarrow \mathcal{S}$. As a result of this transition, a reward is obtained by an agent $i$ at time-step $t$ and it is described as a function of the state and action, i.e. $r_t^{(i)}: \mathcal{S}\times \mathcal{A} \rightarrow \mathbb{R}$\footnote{In the Dec-POMDP, these sets are defined for each agent separately}. Each agent $i$ receives an observation related to the state as $o_t^{(i)}: \mathcal{S} \rightarrow \mathcal{O}$. The DRL framework is summarized in Fig. \ref{fig:drl_framework} where each agent $i$ interacts with the multicast forwarding simulation environment that we have developed on python, `pymanet'.The action value function $Q^\pi$ at time-step $t$ is defined as the future expected cumulative reward under the policy starting from state $s_t$ when the agent executes the action $a_t$ and defined over the time horizon $T$ by
\begin{equation}\label{action-value}
\begin{split}
    Q^\pi(s_t,a_t) &:= \mathbb{E}_t^\pi\left[\sum_{l=0}^T \gamma^l r_{t+l}\right] \\ &=  \mathbb{E}_t^{\pi}[r_t+\gamma Q^\pi(s_{t+1},\pi(s_{t+1}))],
\end{split}
\end{equation}
where the expectation $\mathbb{E}_t^\pi$ is over samples of $r_t \sim \mathcal R(s_t,a_t)$ and $s_{t+1} \sim \mathcal T(s_t,a_t)$. The action-value function can be recursively calculated as (\ref{action-value}) where it describes the well-known Bellman equation \cite{sutton1988learning}.

\subsection{DRL Algorithm}
\begin{figure}[ht]
\centering
\includegraphics[width=0.45\textwidth]{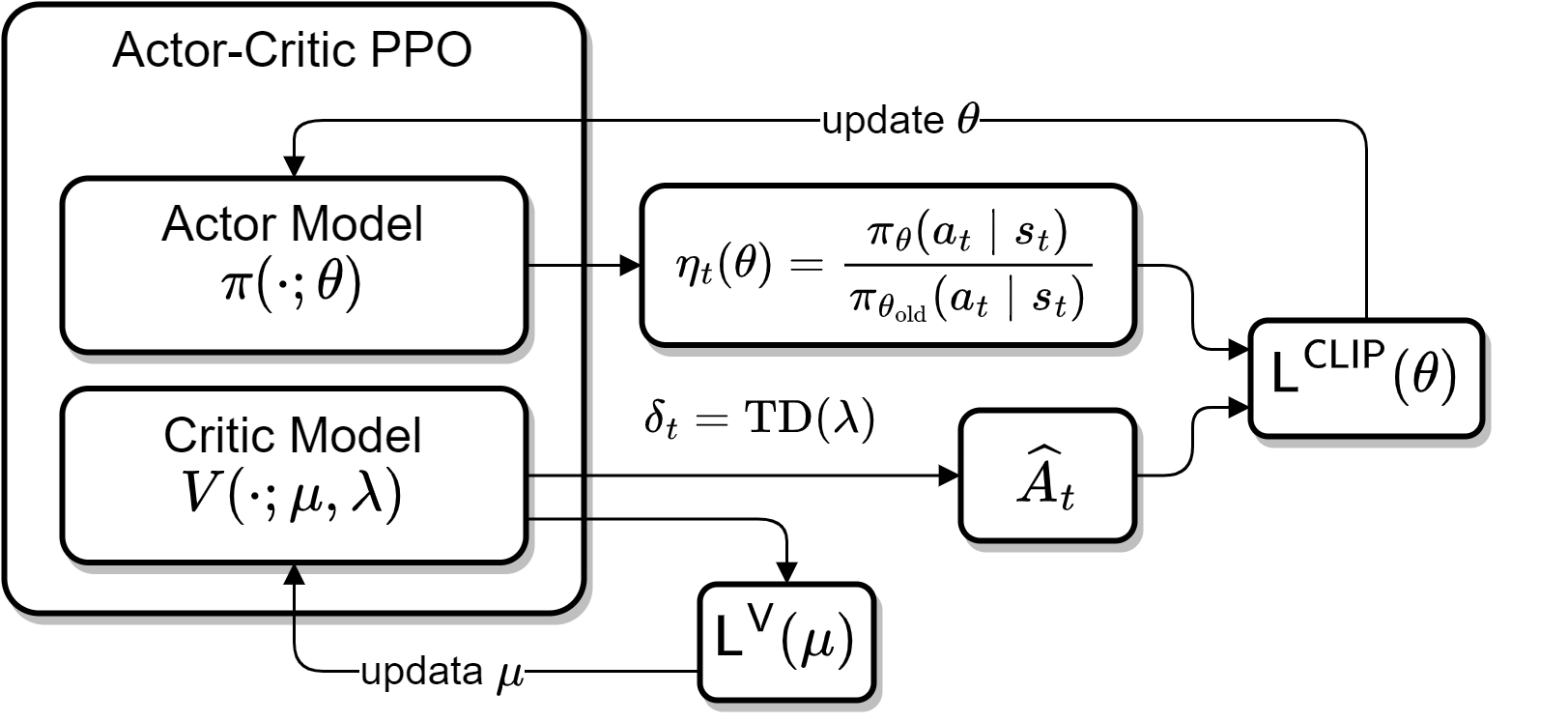}
\caption{\label{fig:ppo} Actor and critic models trained separately in PPO algorithm.}
\end{figure}

Various types of DRL algorithms are introduced in the literature including policy gradients, value-based, actor-critic, and model-based. In this work, we have deployed a popular policy gradient algorithm, i.e. weight-sharing proximal policy optimization (PPO) \cite{schulman2017proximal} which is known for its stability and effectiveness in training robust policies. The main components of the PPO is illustrated in Fig. \ref{fig:ppo}. PPO achieves this balance by employing two main techniques: clipping and surrogate objective. The surrogate objective function used in PPO is given by 
\begin{equation}
\mathsf{L^{CLIP}}(\theta) = \mathbb{E}_t\!\left[\min\!\left(\eta_{\theta_t}\widehat{A}_t(\lambda), \clip\!\left(\eta_{\theta_t},1-\epsilon, 1+\epsilon\right)\!\widehat{A}_t\right)\right] 
\end{equation}
where $\eta_\theta = \frac{\pi_\theta (a_t|s_t)}{\pi_{\theta_\text{old}}(a_t|s_t)}$ and $\theta_{\text{old}}$ is the vector of policy parameters before the update. $\widehat{A}_t(\lambda)$ is the Advantage function using generalized advantage estimate (GAE) with $\lambda$ parameter \cite{sutton2018reinforcement}. 
The clipping mechanism inside the surrogate objection of the PPO constrains the policy update by limiting the change in the policy's probability ratio between consecutive iterations. This prevents large policy updates that may lead to unstable learning. As a result, PPO maximizes the surrogate objective while still respecting the clipping constraint. This approach ensures that the policy update is beneficial while preventing it from deviating too far from the previous policy. By optimizing the surrogate objective function iteratively and updating the policy, PPO gradually improves the agent's performance in the given task. It has been successfully applied to a wide range of tasks, including game-playing, robotics, simulated environments, and even in human feedback training of the chatGPT \cite{ouyang2022training}.

\subsection{Action and Observation Space}
It is critical to design the observation and action spaces in a way that enables our DRL algorithm to train successfully. Since this network model is a multi-agent environment, the observations and actions are defined per nodes or agents. The observations $o_t^{(i)}$ is the observation at node $i$ and at the time $t$. The observation needs to 1) contain only the locally available information of the node's state $i$ 2) maintain ordering information of the nodes 3) be simple and as small as possible to facilitate training of the neural network policy and hesitate from unnecessary excess computational requirements. 4) comply with the MDP (or Dec-POMDP) framework meaning that the actions will define the transition from the current state $s_t$ to the next state $s_{t+1}$. In particular, it cannot be defined as dependent on the packet source (previous hop), as the next incoming packet is independent of the actions and current state of the node. For each node $i$ we use the node ordering following a circular permutation described as 
\begin{equation}
    \psi_i = \left[i, i+1,\ldots, n, 1,  \ldots, i-1\right]
\end{equation}
This is important to ensure DeepMPR policy able to break ties between the nodes in their MPR selection similar to previous heuristic techniques such as S-MPR. We assume that the local link (or adjacency) matrix $\widehat{\mathbf{L}}_i$ at node $i$ is ordered according to the permutation $\psi_i$. The observation is defined by 
\begin{equation}
    o_t^{(i)} = \left[\widehat{\mathbf{L}}_{i,1}, \widehat{\mathbf{L}}_{i,1}^T, \mathbf q_i\right]
\end{equation} 
where it is vector stack of the first row $\widehat{\mathbf{L}}_{i,1}$ and column $\widehat{\mathbf{L}}_{i,1}^T$ of the local link (adjacency) matrix $\widehat{\mathbf{L}}_i$, and the locally available information on queue length at node $i$ and its neighbors. 

The decisions or actions are made as the forwarding decision for the received packet incoming from a hop $h$ to whether or not forward (rebroadcast) into the network. Since due to the Dec-POMDP framework, it is preferred to not include the packet's source as part of the observation, we consider a multi-dimensional action space $a_t \in \{0,1\}^m$ where $m$ is the number of the selected subset of neighbors of $i$ that can send the packet directly to $i$. is given. The observations for each node $o_t^{(i)}$ is extracted from the environment based on the local adjacency information and node embedding. The RL framework is summarized in Fig. \ref{fig:drl_framework}. The actions are chosen by the DNN multicast forwarding policy, DeepMPR, $\pi_\theta(o_i,a_i;h)$ for each packet received from the previous hop $h$ to added to the node's queue and flood the packet or not.

\subsection{Reward}
We have used the instantaneous goodput rate to the neighbors as the local reward component, $r_i$ for each agent $i$. The total reward is a tuned linear combination of the local rewards at the nodes in the network.

\subsection{Centralized Training, Decentralized Execution}
\begin{figure}[t]
\centering
\includegraphics[width=8.4cm]{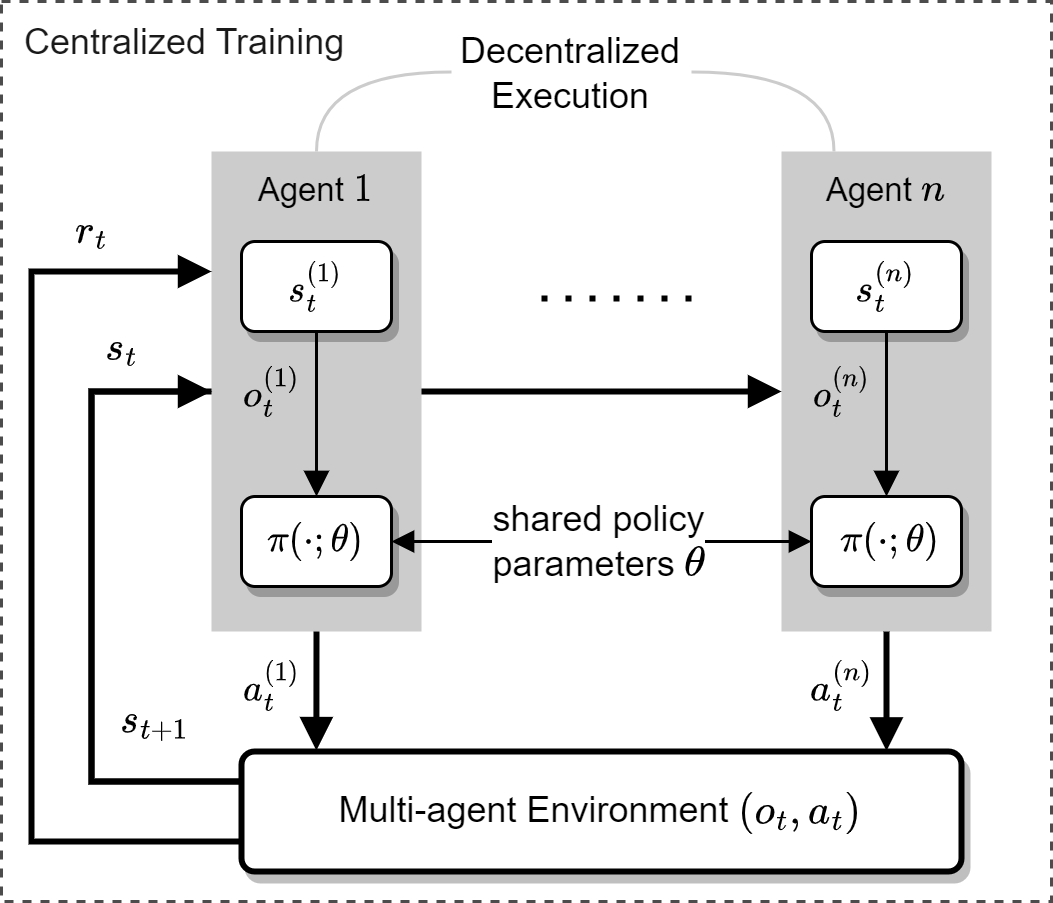}
\caption{centralized training and decentralized execution shown with agents' shared policy. Each agent $i$ uses the shared policy $\pi_\theta$ individually to find its own action $a_t^{(i)}$ based on its own local observations $o_t^{(i)}$.}
\label{decentralized_exec_centralized_train}
\end{figure}

\begin{figure}[ht]
  \centering
  \includegraphics[width=0.44\textwidth]{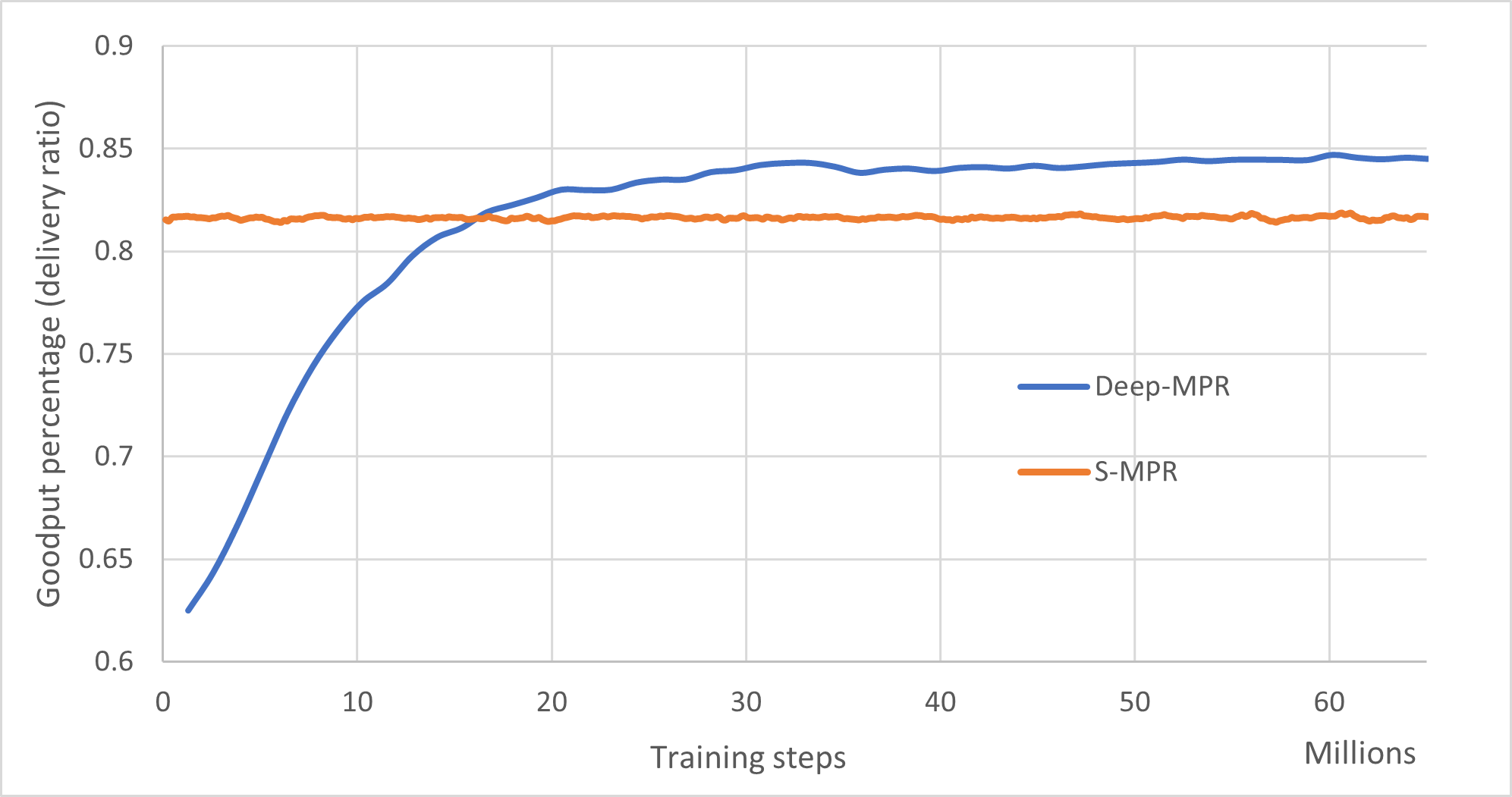}
  \caption{Evaluation results for the delivery ratio (goodput percentage) during training of Deep-MPR compared with the S-MPR versus training.}
  \label{fig:goodput_training}
\end{figure}

\begin{figure}[ht]
  \centering
  \includegraphics[width=0.44\textwidth]{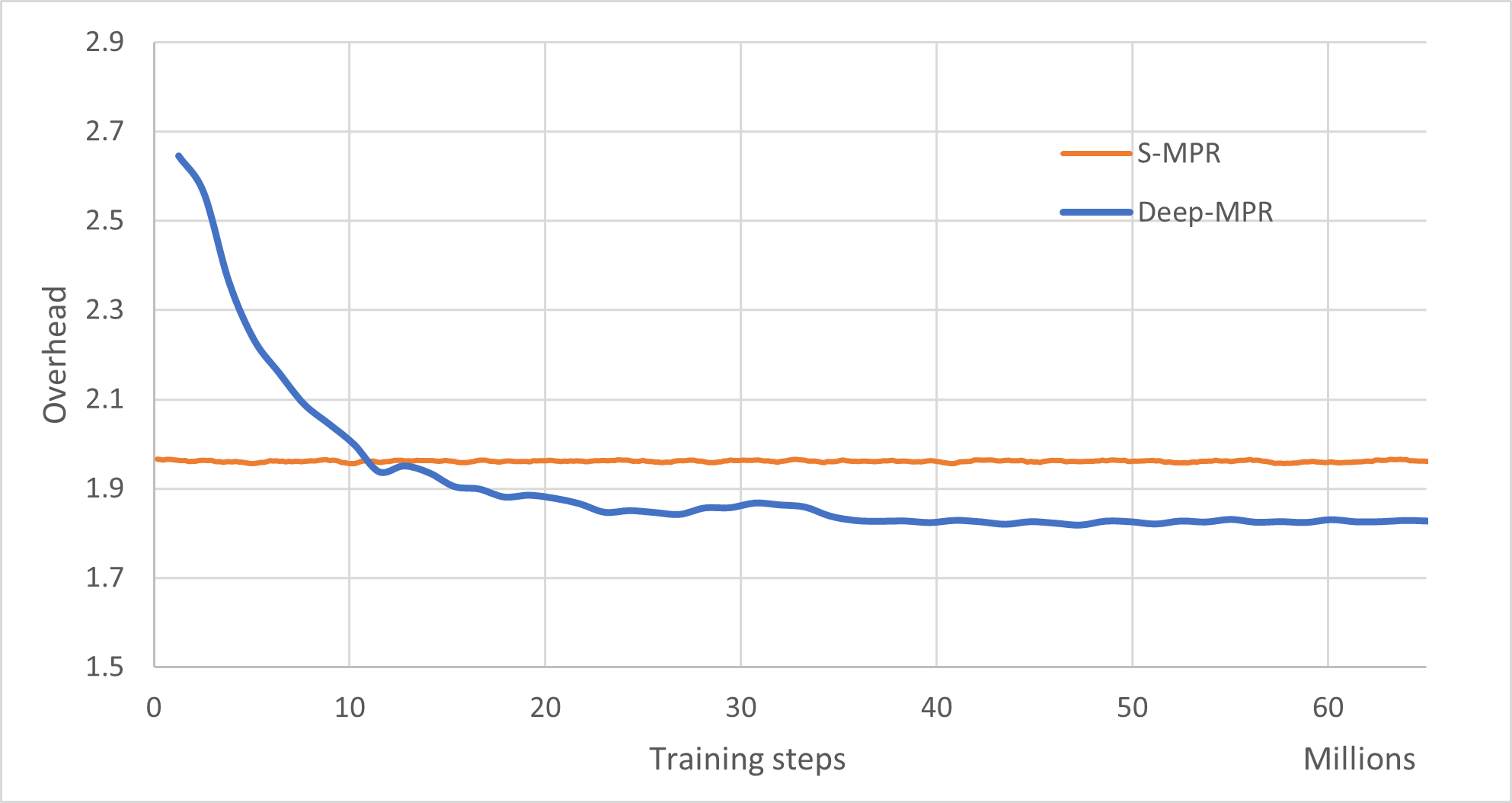}
  \caption{Evaluation results for the overhead ratio (throughput to goodput ratio) during training of Deep-MPR compared with the S-MPR versus training.}
  \label{fig:overhead_training}
\end{figure}

In multi-agent reinforcement learning, each agent can have its policy while sharing the environment with other agents. The routing in MANETs necessitates the learning of decentralized policies, which relies only on the local action-observation history of each agent. Decentralized policies also avoid the exponentially growing joint action space with the number of agents, and therefore more practical and faster to converge in training. Fortunately, decentralized policies can be learned in a centralized fashion specially in a simulated or controlled environment \cite{oliehoek2016concise}. We also use a common strategy to share the policy parameters between agents. i.e. parameter sharing \cite{terry2020parameter}. The parameter sharing is also the enabler for the robustness and scalability of our solution where any added node can execute using the same policy parameters without disruption to the deployment. These approaches are summarized in Fig. \ref{decentralized_exec_centralized_train}.
The results of the training are summarized in Fig. \ref{fig:goodput_training} and Fig. \ref{fig:overhead_training} where it shows Deep-MPR outperforms S-MPR after 15 million steps of the training which includes around 10,000 episodes of training (each episode about 100 seconds of network simulated time). We have deployed our python-based simulation environment, pymanet, and used ray distributed computing platform for parallel training \cite{moritz2018ray, liang2018rllib} which consumed only one day of training on 128 CPUs.

% Formulating MPR selection algorithm based on the adjacency matrix $A = [a_{ij}]$
% $A^2 = \left[a^{(2)}_{ij}\right]$ where $a^2_{ij} = \sum_k a_{ik} a_{kj}$  and it is non-zero when there is a path from $i$ to $j$  of length 2. 
% Current computing nodes is $v$
% $N_1 = \{j : a_{vj} = 1\}$
% $N_2 = \{k \neq v : a^{(2)}_{vk} > 0 , a_{vk} = 0\}$
% \begin{figure}[ht]
%   \centering
%   \includegraphics[width=0.45\textwidth]{robustness_to_mobility_1kbps.png}
%   \caption{Robustness to congestion, 1 source, 25 receivers, 1 m/s max speed, the incoming data flow is increased}
%   \label{fig:your_figure_label}
% \end{figure}

% \begin{figure}[ht]
%   \centering
%   \includegraphics[width=0.45\textwidth]{robustness_to_congestion_4src.png}
%   \caption{Robustness to congestion, 4 source, 25 receivers, 1 m/s max speed, the incoming data flow is increased}
%   \label{fig:your_figure_label}
% \end{figure}

% ===========================================================
\section{Numerical Results}
\begin{figure}[ht]
  \centering
  \includegraphics[width=0.44\textwidth]{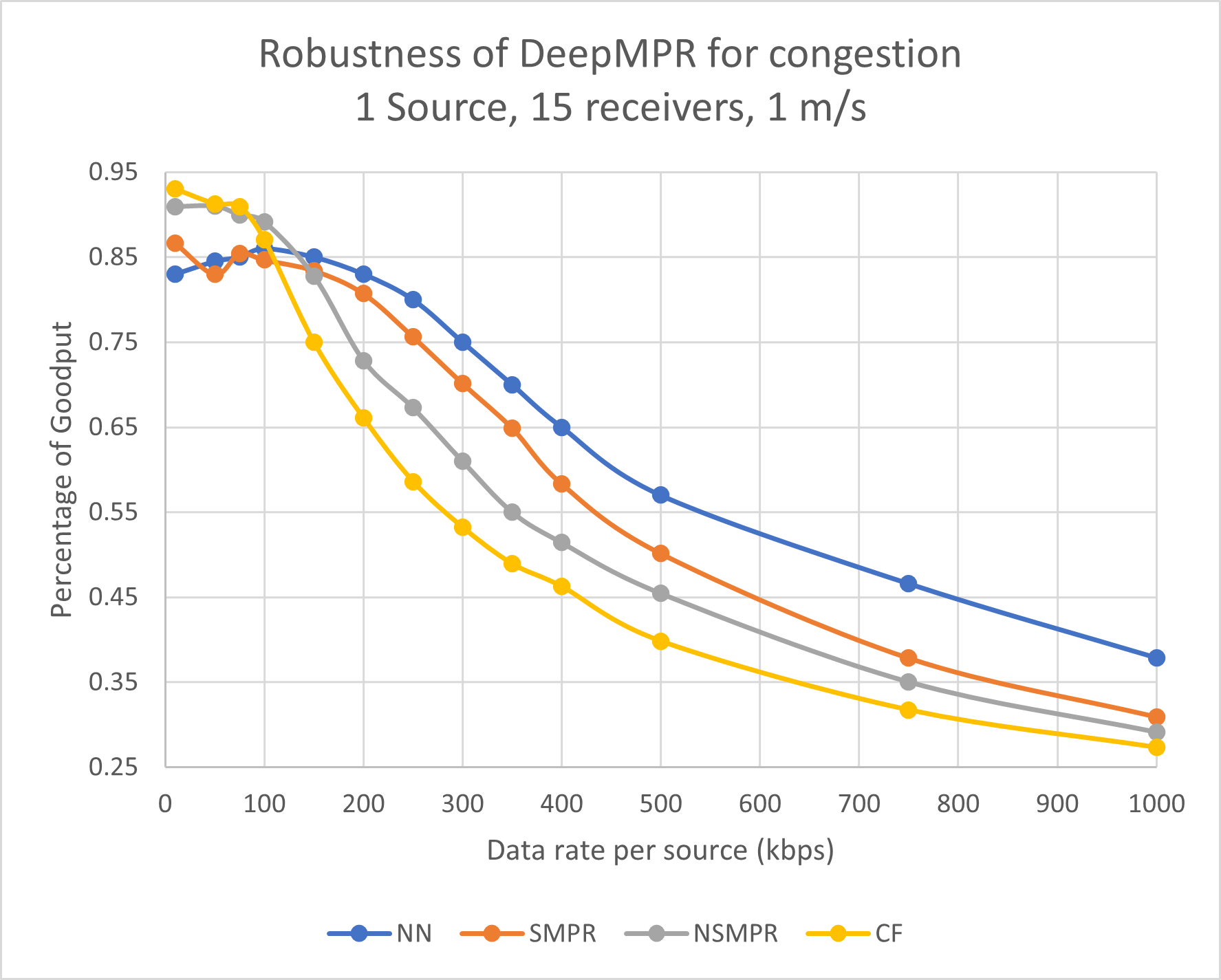}
  \caption{Robustness to congestion, 1 source, 25 receivers, 1 m/s max speed, the incoming data flow is increased}
  \label{fig:robustness_to_congestion}
\end{figure}

For the test and evaluation of our training results, we consider scenarios with the 15 and 25 nodes moving with the Gauss-Markov MANET mobility model within the square area of lengths 550 and 700 meters. The nodes movement will mirror when the hit the borders of this area. Node mean velocity is varying from 1 m/s to 20 m/s. Initial node placement and motion vectors are randomly generated. The radio propagation model is dependent on the range. The maximum range is 250 meters, but the link loss will gradually increase from 200 meters of the range until it disconnects at 250 meters. The OLSR \texttt{HELLO} messages are sent at every $0.5 \pm 0.25$ seconds. The multicasting data flow is generated from the source nodes with a random uniform distribution. Multicast packet data is sent in 256-byte UDP datagrams with a time-to-live of 255 hops. Considering the typical link data rate of 1 Mbps for 802.11 standard, each packet will have slightly larger than 2ms over the air transmission. Our results during training and evaluations show improvements in delivery ratio particularly when the network is congested with the higher rate of the multicast flows. We have evaluated the performance of the proposed policy over the delivery ratio as shown in Fig. \ref{fig:robustness_to_congestion}. The percentage of the goodput to the total deliverable goodput shows up to \%10 higher goodput when the network is congested by the multicast packet shows significant enhancement in network overhead. 

% ===========================================================
\section{Conclusions}
We have developed a DRL-based framework that aims to enhance the MPR subset selection in a MPR-based multicast routing protocol, without the need for MPR announcement. Our framework focuses on designing a DNN policy for efficient multicast routing. The results of our study demonstrate a substantial improvement in multicast packet delivery ratio, especially in congested network scenarios. This improvement is achieved through intelligent forwarding decisions made by our Deep-MPR multicast forwarding policy.

\section{Acknowledgements}
Research reported in this publication was supported in part
by the Office of the Naval Research under the contract N00014-
19-C-1037. The content is alone the responsibility of the
authors and does not necessarily represent the official views of
the Office of Naval Research. The authors would like to thank
Dr. Santanu Das (ONR Program Manager) for his support and
encouragement.

% -------------------------------------------------------------

\bibliographystyle{IEEEtran}
\bibliography{DeepMPR.bib}
\end{document}